# Electron drift velocity control in GaAs-in-Al$_2$O$_3$ quantum wire transistor structure due to the electron scattering rate alteration

*A.V. Borzdov, D.V. Pozdnyakov\*, V.M. Borzdov*

Radiophysics and Electronics Department, Belarusian State University,
Nezavisimosty av.4, 220030 Minsk, Belarus

\* Corresponding author: e-mail pozdnyakov@bsu.by

Electron transport in the transistor structure based on thin undoped GaAs-in-Al$_2$O$_3$ quantum wire is simulated by ensemble Monte-Carlo method taking into account electron scattering by the phonons and surface roughness. The influence of surface roughness height on electron drift velocity at 77 and 300 K is investigated for the values of longitudinal electric field strength of $10^4$ and $10^5$ V/m. A possibility of electron drift velocity control due to variation of the bias applied to the gates, which results in the electron scattering rate alteration, is ascertained.

**1 Introduction** In the past few decades a great attention has been paid to the GaAs-based field effect transistors with one-dimensional electron gas (1 DEG) or, in other words, the quantum wire transistors [1, 2]. The interest in investigation of the quantum wire transistors is caused by a further shrinking of the dimensions of integrated circuit elements which reveals the need in comprehensive study of electron transport properties in the channels of such devices.

One of the ways to scale down the dimensions of the field effect transistors and increase their performance is the use of A$^{III}$B$^V$ semiconductors as conductive channels and Al$_2$O$_3$ as the gate dielectric [3, 4]. Using Al$_2$O$_3$ as the gate dielectric allows the gate leakage currents to be reduced and the breakdown voltages to be increased along with narrowing down transistor dimensions.

**2 Simulation model**

**2.1 Calculation of electron states in 1 DEG** In this study we consider a rectangular quantum wire transistor structure consisting of an undoped GaAs channel surrounded by Al$_2$O$_3$ matrix with two gates (see Fig. 1). The electrons in the considered structure are spatially confined within the GaAs channel in $X$ and $Y$ directions which leads to the formation of 1 DEG. To define the electron subband energies and wave functions in 1 DEG one must write down an appropriate system of the Schrodinger and Poisson equations, and solve it self-consistently. While carrying out the calculations we have neglected electron concentration in the transistor channel and supposed that the infinitely high potential barrier for electrons at Al$_2$O$_3$/GaAs interface takes place. Thus the electrostatic potential inside the GaAs channel is determined only by the external electric field caused by the applied gate biases. Also the simulation is performed for the condition of the electric quantum limit, i. e. it is supposed that all the electrons populate only the ground quantum state. The Schrodinger equation in parabolic band approximation for the structure takes the following form

$$\left(-\frac{\hbar^2}{2m^*}\Delta - e\varphi(x,y)\right)\psi_0(x,y) = E_0\psi_0(x,y), \quad (1)$$

where $\hbar$ is the Planck constant, $m^*$ is the electron effective mass, $e$ is the magnitude of electron charge, $\psi_0(x, y)$ is the wave function for electrons in the ground quantum state, $E_0$ is the energy level of the ground quantum state.

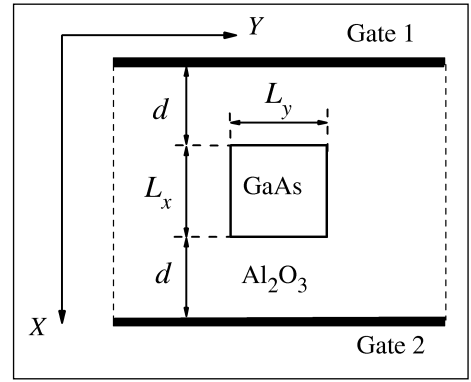

**Figure 1** Cross-section of the considered quantum wire transistor structure with the following dimensions: $L_x = L_y = L = 10$ nm, $d = 10$ nm.

The electrostatic potential $\varphi$ is defined by the corresponding Poisson equation which in our case reduces to the equation

$$\nabla(\varepsilon(x,y)\nabla\varphi(x,y)) = 0, \quad (2)$$

with $\varepsilon(x, y)$ being the static permittivity of the materials.

For simplicity the structure is treated as a plane capacitor neglecting the fringe effects, so that only $X$-component of the electric field is considered. As a result the derivation of solution of Eq. (2) is straightforward. Then the obtained electrostatic potential $\varphi$ is used to find electron energy level and wave function for the ground quantum state via a numerical solution of the two-dimensional Schrodinger equation (1).

**2.2 Calculation of electron scattering rates** The ensemble Monte-Carlo simulation procedure includes the most important electron scattering processes in the thin undoped GaAs quantum wires. In the considered case the dominant scatterers are the localized polar optical and acoustic phonons, and surface roughness [5–10]. The electron scattering rates are calculated including the collisional broadening according to the procedure, described in [11, 12]. As electron concentration is assumed to be negligible in the channel, 1 DEG is treated as nondegenerate, and the Pauli principle is not taken into account in the present study. For the case of electric quantum limit we have adopted the equations for electron scattering rates using the results from Ref. [12]. So, the expression for the acoustic phonon scattering rate can be written as

$$[W_{f,b}]_A(E,\Gamma) = \frac{B_{ac}^2 k_B T \sqrt{2m^*}}{2\hbar^2 v^2 \rho} D(E,\Gamma)$$

$$\times \int_0^{L_x}\int_0^{L_y} |\psi_0(x,y)|^4 dxdy, \quad (3)$$

$$D(E,\Gamma) = \Theta(E)\sqrt{\frac{\Gamma+\sqrt{E^2+\Gamma^2}}{E^2+\Gamma^2}},$$

where $B_{ac}$ is the deformation potential for acoustic phonons, $k_B$ is the Boltzmann constant, $T$ is the temperature, $v$ is the sound velocity in GaAs, $\rho$ is the density of GaAs, $E$ is the electron kinetic energy, $\Gamma$ is the quantity characterizing the collisional broadening of the energy spectrum of charged particles due to all the scattering mechanisms under consideration, $\Theta$ is the unit step function.

The expression for the electron scattering rates for electron–polar optical phonon interactions takes the form

$$[W_{f,b}^{e/a}]_{PO}(E,\Gamma) = \frac{e^2\omega\sqrt{2m^*}}{\hbar L_x L_y}\left(\frac{1}{\varepsilon^\infty}-\frac{1}{\varepsilon^s}\right)\left(n+\frac{1}{2}\pm\frac{1}{2}\right)$$

$$\times D(E\pm\hbar\omega,\Gamma)$$

$$\times \sum_{p=1}^{\infty}\sum_{r=1}^{\infty}\frac{\left|\int_0^{L_x}\int_0^{L_y}|\psi_0(x,y)|^2 \sin(p\pi x/L_x)\sin(r\pi y/L_y)dxdy\right|^2}{(q_{f,b}^{e/a})^2+(p\pi/L_x)^2+(r\pi/L_y)^2}, \quad (4)$$

$$q_f^{e/a} = \frac{\sqrt{2m^*E}}{\hbar}-\frac{\sqrt{2m^*(E\mp\hbar\omega)}}{\hbar},$$

$$q_b^{e/a} = \frac{\sqrt{2m^*E}}{\hbar}+\frac{\sqrt{2m^*(E\mp\hbar\omega)}}{\hbar}.$$

Here $\omega$ is the cyclic frequency of a polar optical phonon; $\varepsilon^\infty$ and $\varepsilon^s$ are the optical and static permittivities of GaAs, respectively; $n$ is the Bose–Einstein distribution function; $L_x$ and $L_y$ are the height and the width of the GaAs channel of the transistor structure, respectively. The superscript "e/a" denotes the emission/absorption of the phonon, and the subscripts "f" and "b" indicate the forward and backward scattering, respectively.

Finally the rate of electron scattering from surface roughness in the GaAs transistor structure is given by the following expression

$$[W_{f,b}]_{SR}(E,\Gamma) = \frac{1}{4}\left(\left(\frac{\partial E_0}{\partial L_x}\right)_1^2+\left(\frac{\partial E_0}{\partial L_x}\right)_2^2\right.$$

$$\left.+\left(\frac{\partial E_0}{\partial L_y}\right)_1^2+\left(\frac{\partial E_0}{\partial L_y}\right)_2^2\right)\frac{\sqrt{\pi}\Delta^2\Lambda\sqrt{2m^*}}{2\hbar^2}$$

$$\times\frac{D(E,\Gamma)}{1+\pi\Lambda^2 k(E)^2((E\mp E+\Gamma)/2)}, \quad (5)$$

where $k(E)=\sqrt{2m^*E}/\hbar$, $\Lambda$ is the roughness correlation length, and $\Delta^2 = \delta^2 g(L^{-1}\Lambda)$ ($\Delta\leq\delta$) [13]. Here $\Delta$ is the amplitude of the roughness [9], $\delta$ is the root-mean-square height of the amplitude of the roughness (the surface roughness height), characterizing the deviation of the surface from a plain [13]. The partial derivatives of the energy level of the ground quantum state $E_0$ with respect to the transverse sizes of the GaAs region $L_x$ and $L_y$ determine the change in the energy when each wall of the quantum well deviates from the plane (subscripts 1 and 2 denote each scattering surface for $X$ and $Y$ directions) and are calculated numerically during the solution of the Schrodinger equation (1).

After solving the system of equations (1) and (2), we can calculate the electron scattering rates from relationships (3)–(5) using the following equation [11, 12]

$$\Gamma = \frac{\hbar}{2}\sum_{\eta=f,b}\left([W_\eta]_A+[W_\eta^e]_{PO}+[W_\eta^a]_{PO}+[W_\eta]_{SR}\right). \quad (6)$$

Equation (6) must be solved numerically for all values of the electron kinetic energy $E$ from the energy range under consideration.

**2.3 Monte-Carlo simulation procedure** Electron transport in the quantum wire structure is simulated by means of the ensemble Monte-Carlo method using the ensemble of $10^6$ particles. The general theory behind the application of the Monte-Carlo method to semiconductor device simulation may be found elsewhere [14]. We should refer the readers to Refs. [15–18] for details of realization of the Monte-Carlo algorithms in the structures with 1 DEG.

In the Monte-Carlo simulation procedure the electron momentum space is divided into cells and the distribution function is defined over the grid in the momentum space by counting the number of electrons in each cell. The simulation procedure starts supposing maxwellian distribution function, and electron ensemble is propagated in time until the steady drift is set up. The electron distribution function $f(p)$, where $p$ is the electron momentum, is updated at every time step and normalized so that the integral of $f(p)$ over the simulated electron momentum interval is unity. Thus $f(p)$ means a probability for an electron to have a momentum value in interval $(p, p+dp)$.

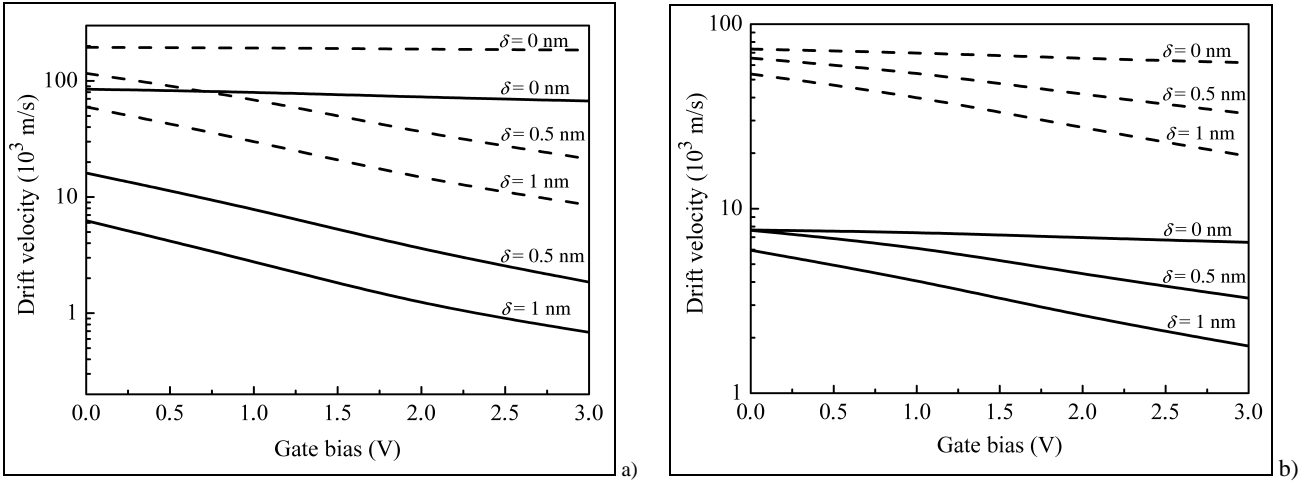

**Figure 2** Dependencies of electron drift velocity on the gate bias calculated at a) $T = 77$ K and b) $T = 300$ K for different values of the surface roughness height $\delta$. The solid curves correspond to the case when the longitudinal electric field strength is equal to $10^4$ V/m, and the dash curves – $10^5$ V/m.

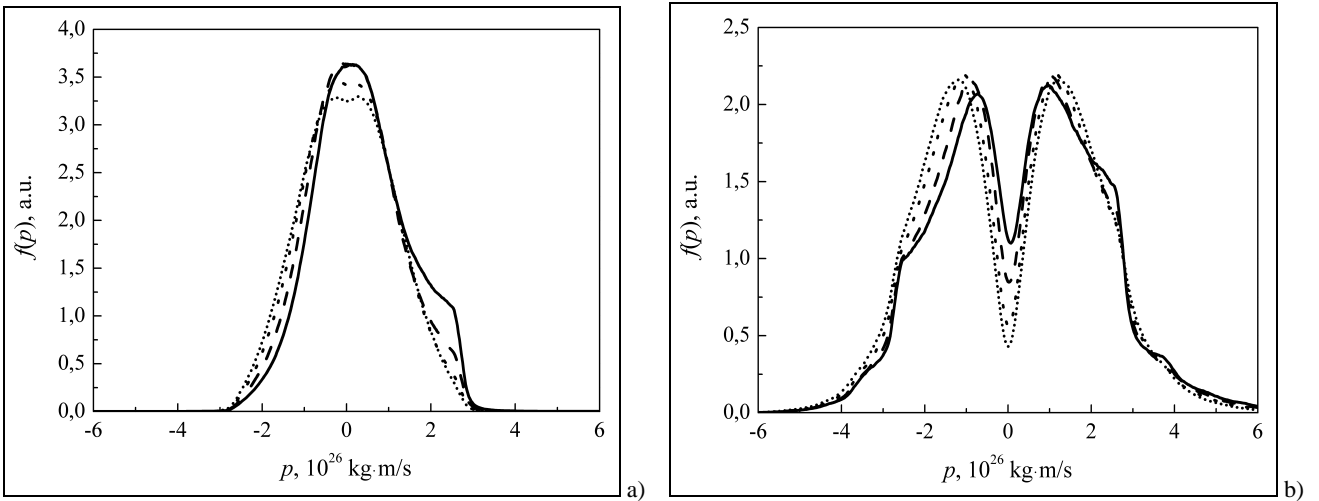

**Figure 3** Electron distribution functions for different values of the gate bias $V_G$ at a) $T = 77$ K and b) $T = 300$ K. Solid curves – $V_G = 0$ V, dash curves – $V_G = 1$ V, dot curves – $V_G = 2$ V, short-dot curves – $V_G = 3$ V. The longitudinal electric field strength is equal to $10^5$ V/m, and the surface roughness height $\delta$ is equal to 1 nm.

**2.4 Results and discussion** To provide the highest possible performance of a field effect transistor it is essential to reduce electron scattering rates in the transistor structure and thus to achieve a higher electron drift velocity. Creation of high quality GaAs/Al$_2$O$_3$ boundaries can minimize the effect of a surface roughness scattering on electron drift velocity. However, we propose to utilize the surface roughness scattering for the drift velocity control in the thin GaAs quantum wires. In Refs. [12, 19] it is shown that in GaAs-in-AlAs quantum wire transistor structure the surface roughness scattering rate is sufficiently influenced by the gate bias. The latter effect occurs due to the rise of the magnitude of the partial derivatives of subband energy level with regard to the well width in the direction of the electric field strength. Taking into account a rather strong dependence of surface roughness scattering rates on the transverse electric field strength, one could utilize this feature of the quantum wire transistor structure to control the electron drift velocity in the channel by applying gate bias to one of the gates with respect to another one (let us suppose that the bias is applied to one of the gates, say gate 1, while the bias of the gate 2 is set to zero).

In Fig. 2 the electron steady-state drift velocity versus gate bias is presented. The electron drift velocity is calculated for several values of the surface roughness height $\delta$ with constant value of the roughness correlation length $\Lambda = 6$ nm. As it can be seen for $\delta = 0$ (the surface roughness scattering does not take place) the gate bias effect on the drift velocity is insignificant at both 77 and 300 K. A slight decrease of the drift velocity is observed due to the growth of electron-phonon interaction caused by the electron wave function deformation in the presence of transverse electric field [12]. For $\delta > 0$ the presence of surface roughness scattering begins to play a decisive role in the decrease of the drift velocity with the growth of the gate bias. The influence of the surface roughness scattering is pronounced enough for both low and high temperatures. The surface roughness scattering rates do not depend on the temperature explicitly except for the collisional broadening factor [12] while the phonon scattering rates increase and begin to dominate at high temperatures. Nevertheless, if the parameter $\delta$ is large enough the effect of the surface roughness scattering is obvious even at 300 K. Such a phenomenon can be exploited to control the electron drift velocity in the transistor channel by the application of the gate bias.

The calculations revealed another interesting effect of the gate bias on the electron drift velocity in the presence of surface roughness. Comparing the electron drift velocity dependencies on the gate bias for given values of $\delta > 0$ at 77 and 300 K, one can note that at high biases the drift velocity falls down to lower values at 77 K than at 300 K though the overall scattering rates are higher at 300 K. The latter may seem controversial, but the behaviour of the drift velocity may be explained while analyzing the electron distribution functions.

The electron steady-state distribution functions $f(p)$ for 77 and 300 K at the longitudinal electric field strength equal to $10^5$ V/m and different gate biases are presented in Fig. 3. As it can be seen from the figures, there is a substantial dip in the distribution functions in the region near the zero momentum value at 300 K whereas there is no such a dip at 77 K. This effect must be related to the fact that the electron scattering rate with absorption of polar optical phonon is much higher at 300 K than at 77 K. Also one must take into account that for large values of $\delta$ the surface roughness scattering is dominant for low electron kinetic energies and decreases with the increase of the energy. Then electrons are efficiently "pumped" out from the region with low kinetic energies and high surface roughness scattering rates to the region with high energies and low surface roughness scattering rates at high temperatures. Due to such an electron pumping it occurs that the majority of electrons occupy the states in momentum space with larger momentum relaxation times at 300 K than at 77 K.

**3 Conclusion** The results of Monte-Carlo simulation of electron drift in the thin gated GaAs-in-Al$_2$O$_3$ quantum wire are presented in this study. It is shown that due to the peculiarities of surface roughness scattering processes in quantum wire structures like that it is possible to modulate the value of electron drift velocity in the channel efficiently via the application of an external bias to one of the gates with respect to another one. In addition we want to note that our future purpose is a further research of influence of some high-order quantum effects [20] on the electron transport in the GaAs-in-Al$_2$O$_3$ quantum wire transistor structures.